\documentclass{article}

\usepackage{arxiv}
\usepackage[utf8]{inputenc} 
\usepackage[T1]{fontenc}    
\usepackage{hyperref}       
\usepackage{url}            
\usepackage{booktabs}       
\usepackage{graphicx}
\usepackage{epstopdf}
\usepackage{amsfonts,amsmath,amssymb}       
\usepackage{nicefrac}       
\usepackage{microtype}      
\usepackage{lipsum}		
\usepackage{doi}
\usepackage{siunitx}
\usepackage{textcomp}
\usepackage{xcolor}

\title{Decomposition of Industrial Systems for Energy Efficiency Optimization with OptTopo}

\author{ 
\href{https://orcid.org/0000-0002-7108-5203}{\includegraphics[scale=0.06]{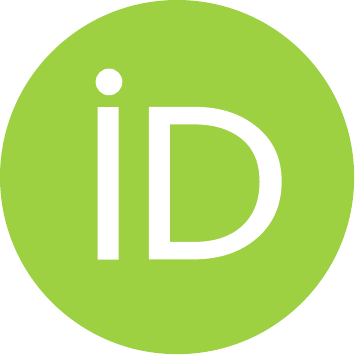}\hspace{1mm}Gregor Thiele}\\
	Department for Process\\
	Automation and Robotics,\\
	Fraunhofer Institute for\\
	Production Systems\\
	 and Design Technology IPK\\
	\texttt{0000-0002-7108-5203} \\
	\And
	\href{https://orcid.org/0000-0002-0562-669X}{\includegraphics[scale=0.06]{orcid.pdf}\hspace{1mm}Theresa Johanni} \\
	Faculty of Electrical \\
Engineering\\
and Computer Science\\
Technical University of Berlin\\
Berlin, Germany \\
	\texttt{0000-0002-0562-669X} \\
\And
	\href{https://orcid.org/0000-0002-0562-669X}{\includegraphics[scale=0.06]{orcid.pdf}\hspace{1mm}David Sommer} \\
	Weierstrass Institute\\
for Applied Analysis\\
and Stochastics\\
Berlin, Germany \\
\texttt{0000-0002-6797-8009} \\
\And
	\href{https://orcid.org/0000-0002-0562-669X}{\includegraphics[scale=0.06]{orcid.pdf}\hspace{1mm}Jörg Krüger} \\
	Institute for Machine Tools\\
and Factory Management\\
Technical University of Berlin\\
Berlin, Germany \\
	\texttt{0000-0001-5138-0793} \\
}

\renewcommand{\headeright}{}
\renewcommand{\undertitle}{}
\renewcommand{\shorttitle}{Decomposition of Industrial Systems for Energy Efficiency Optimization with OptTopo}

\hypersetup{
pdftitle={A template for the arxiv style},
pdfsubject={q-bio.NC, q-bio.QM},
pdfauthor={David S.~Hippocampus, Elias D.~Striatum},
pdfkeywords={First keyword, Second keyword, More},
}

\begin{document}
\maketitle

\begin{abstract}
	The operation of industrial facilities is a broad field for optimization. Industrial plants are often a) composed of several components, b) linked using network technology, c) physically interconnected and d) complex regarding the effect of set-points and operating points in every entity. This leads to the possibility of overall optimization but also to a high complexity of the emerging optimization problems. The decomposition of complex systems allows the modeling of individual models which can be structured according to the physical topology. A method for energy performance indicators (EnPI) helps to formulate an optimization problem. The optimization algorithm OptTopo achieves efficient set-points by traversing a graph representation of the overall system.
\end{abstract}


\section{Introduction}

Using energy as efficiently as possible is a key challenge at this time. For the manufacturing sector this gives rise to investment in more machinery as well as approaches to use existing facilities more efficiently. This paper contributes to the last, specifically to the set-point optimization of complex systems by means of decomposition. We use the physical topology information about the plant structure in order to model components separately and organize these potentially heterogeneous models in a graph. We applied a method for energy performance indicators (EnPI) designed by ÖKOTEC in order to formulate an optimization problem \cite{ThieleGregor2019Ffee}. This can be solved recursively by the optimization algorithm OptTopo which traverses the graph containing all component models. A draft of the approach was first presented in 2020 \cite{optTopoCodit2020}.
Further details about performance measures are given in \cite{optTopoCodit2022}.

\section{Related Work}

\paragraph{Cascade control for coupled HVAC systems} Komareji et al. consider a system which consists of several heat exchangers whose set temperatures are regulated by presetting the volume flows of air and water \cite{Komareji}. A primary, a secondary and a tertiary circuit are coupled, the first transition being realized by an air-air heat exchanger, the second by an air-water heat exchanger. For clarity, the relationships are modeled as static polynomials. Since polynomial functions up to the third degree are used, it is a nonlinear problem which is potentially nonconvex. The numerical calculation is not detailed in the article, but the decomposition of interconnected utilities and the modeling with static functions is described as an example. With Komareji \cite{Komareji,Komareji2} the optimization is done by an objective function from models per component. Thus, the optimization is done for coupled systems, but rather not along the topology, because all manipulated variables are changed in the overall model. Only continuous variables occur and a non-convex model is explicitly assumed.

\paragraph{Optimization of Machine Tools Using Graph Models} Working in the ECOMATION research group, Eberspächer and Schlechtendahl \cite{EberspaecheraPhilipp2013RERo,Schlechtendahl.2016} describe a graph of states that is also developed from a consideration of the components, leading to optimization of coupled systems. The optimization is a set of rules to navigate along this state machine. Any continuous and integer values can be stored in the nodes of this state machine but the optimization is still the selection of solutions from a finite set. This means there is no need to distinguish between integer and real values here, and in particular convexity is not a category in this context. This approach is related to OptTopo but does not necessarily consider Kirchhoff's law for nodes with respect to energy flows because machine tools do not have the chained energy transformation like cooling plants do.

\paragraph{A Multiplexed Real-time Optimization (MRtOpt)} Asad et al. presented MRtOpt for air-conditioning in an office building \cite{AsadHussainSyed2017Mroo,Asad2016}. Their approach represents a supervisory cascade for several locally controlled units in order to reduce the total consumption. The authors vividly present the mutual influence of the involved subsystems, which emphasizes the complexity of the resulting optimization problem. An \textit{interior point method} is used to solve the posed optimization problem.

Testing was done by means of a simulation with the established tool TRNSYS. The authors point out potentially undesirable deviations in the presence of abrupt changes in setpoints \cite{Asad2016} and design separate models for predicting these effects, both for the actuated control loop and for neighboring entities physically affected by the change. The models are determined by \textit{sub-space system identification}. An adaptive variant of the method was published in 2019  \cite{AsadHussainSyed2019Amfr}.

The MRtOpt according to Asad et al. refers to a system of several components, thus optimizing coupled systems, but as a joint objective function without considering the topology. Polynomials are used, making the models transferable and transparent. A convex and continuous problem is assumed.

\paragraph{Extremum seeking control (ESC)} In a collaboration between Whirlpool Cooperation, Johnson Controls Inc. and the University of Texas, Baojie et al. investigated the optimization of a refrigeration system consisting of two chillers operating in parallel, a cooling tower and associated water circuits \cite{MuBaojie2017Rooa}. In optimization of building services, the authors first distinguish rule-based, model-based and model-free optimization. ESC is used, where the gradient is estimated along the way. The system is stimulated by a peripheral signal (in this case a sinusoidal oscillation). Due to the provoked variation around the operating point, the gradient can be estimated robustly. This is integrated into the control so that the operating point is shifted in real time in the direction of lower energy consumption.

Addressed manipulated variables are the speed values of the chilled water pumps, the ventilation of the cooling tower and the supply temperatures of the chillers. The authors assume the input and output dynamics each as a LTI system, thus encapsulating any nonlinearities as a static model and assuming the convexity of the optimization problem. The control itself is performed by a PI controller. When the manipulated variable is constrained, the integrator may ineffectively assume an unrealistically high value, but without reaching the distance accordingly. This wind-up effect poses the risk that the integrator value cannot be decayed fast enough when the motion is in the opposite direction. Ying et al. \cite{esc_constrained} present an anti-windup solution for ESC controls.

Complementing the presented method, Salsbury et al. published an automatic selection of the considered input variables \cite{SalsburyESC2022}. The starting point is again a \textit{dipher signal} composed of several oscillations to stimulate the system dynamics. Here, for inputs, it is checked concurrently whether there is no coupling or a weak or strong coupling. This is done by means of singular value decomposition. Thus, only those couplings with large singular values are considered to keep the complexity of the optimization problem low. This ESC approach was tested on a Modelica simulation in \cite{MuBaojie2017Rooa} with anti-windup control and in \cite{SalsburyESC2022} with automatic input selection. 

The extremum seeking control after Baojie u. Salsburg forms analogously to Komareji \cite{Komareji,Komareji2} a functional from separate terms for each component. In contrast to Komareji, however, no prefabricated models are used. An empirical estimation of the gradient allows the model-free descent on the functional, which is assumed to be convex and continuous for this purpose.

\paragraph{Multi-Agent Reinforcement Learning (MARL) in machine control} This approach was presented by Bakakeu in a cooperation with SIEMENS Digital Factory division \cite{BakakeuJupiter2020MRLf,Bakakeu2021}. With this algorithm, several machines are to run in such a coordinated manner that the joint energy consumption, taking into account self-generation and volatile energy prices, enables an always up-to-date optimum. For this purpose, several agents are used, for which, however, the Markov condition can no longer be assumed for the mutual influence. 

The learning of the agents follows the scheme of the autocurricula \cite{DBLP:journals/corr/abs-1903-00742}, so that challenges are solved cooperatively and competitively. The agents learn the behavior of the surrounding machines along with the strategies of the agents behind them. The implementation, like Panten's, uses Proximal Policy Optimization (PPO). The authors describe how the machines shut themselves down when overload threatens, but act swiftly on changes in the situation and compete for supply phases that become available.
Bakakeu's Multi-agent reinforcement Learning takes into account several components of a system, each with autonomous agents which, however, partly collaboratively pursue common goals. The models are created by learning, so they are not directly transferable and also not transparent. MARL can provide individual agents with different initial values, but this is not explicitly discussed in the paper. Integer values and binary states can easily be included in the state spaces of MARL.

\paragraph{Deep Reinforcement Learning (DRL)}
Panten submitted as a dissertation in 2019 a study on the use of DRL to increase energy efficiency in utility engineering \cite{panten_dissertation}. Panten argues the necessity of machine learning methods with the increase in complexity of optimization tasks due to both the interconnection of many subsystems and the consideration of dynamic memory effects. For the latter, prediction functions are provided. The scalability of the optimization facilitates a hierarchical concept, which solves subproblems locally on a lower level, whereby only a fraction of the variables enter a higher-level optimization problem.

Panten builds a simulation of a complex system that includes a cooling tower, chillers, pumps, condensing boilers, immersion heaters, solar panels, and other elements, relying on FMI to organize the models created in Modelica. The components each have local control, such as by hysteresis or PI controllers. The higher-level system behavior is learned by a deep neural network (DNN) using PPO. Reinforcement learning now optimizes continuous parameters on this system knowledge, which are passed to the lower-level control. The communication between the entities is done via OPC UA. The entire approach is validated on simulation data using a functional model implemented in Python. \cite{pantenCirpAnnals}

Panten's approach is similar to the approach according to Bakakeu. Here, too, coupled systems are considered, but in Panten's case the superimposed optimization takes place without local agents being pronounced for each entity. Thus, one cannot speak of an optimization along the topology here. With the DNN approach, models can neither be interpreted manually nor transferred in a modular way. In this study, the optimization parameters are expressed as real numbers. In general, a generalization to integer and binary attributes should be possible here as well.

\paragraph{Optimization of Parallel Pumps} Wang and Zhao present a special case of using topology where six differently sized pumps interact to achieve the flow and pressure requirements with as little energy as possible \cite{WangXuetao2020ADOM,WANG2020100001}. These operating requirements are assumed to be quasi-static. The pumps can be turned on and off and controlled by flow rate, and the resulting behavior is described with polynomials estimated from measured data.

The pumps operating in parallel each have controllers, which in turn are connected in a network. This results in a network of agents that can be represented as a connected graph. A starting node sends messages to neighboring nodes, which identify themselves as child nodes and also send out messages until all (six in the example) nodes are marked. The nodes evaluate randomly chosen operating points along a uniform distribution. These include normalized speed $\omega$, flow $Q$, and electrical power $P$. The algorithm includes synchronous updates of this aggregation and replaces the value initialized by $P^\star = \infty$ for the total power expended. For a sufficient number of iterations, this algorithm finds the optimal allocation of flow among the six pumps. For the example,  the desired setting could be found with \num{20} iterations in under \SI{3}{\second}. In the experiment, Zhao et al. apparently use the polynomials of the pump characteristics both for the solution algorithm and for validation, which makes it difficult to track. A more advanced publication by the same authors converts the method to asynchronous communication, while preserving the basic principle \cite{WANG2020100001}.

This approach deals with coupled pumps, but considers the topology only to establish a computational order and is limited to a parallel arrangement. Polynomials are deposited, which is why the models are interchangeable and transparent. The consideration of discrete states is conceivable, but not explicitly discussed.

\section{Concept}
\label{sec:concept}
\subsection{General Workflow to Optimize a System Based on Topology}
For the purpose of the decomposition of complex optimization problems we propose an optimization methodology which comprises as complete workflow process the following steps:

\begin{enumerate}
  \item The decomposition of the industrial system to be optimized in independent subsystems together with a definition of energy (or material) flows that connect these subsystems and build their topology (and the formalization of this decomposed system as a graph with nodes and edges);
  \item The generation of models for the in- and outgoing flows of all the subsystems (since they are separated, these models can be of different forms for subsystems of one overall system, like using polynomials together with Neural Networks for different subsystems);
  \item Use the system decomposition ( step 1) together with the models of the subsystems (step 2) and an optimization functional as input for an optimization algorithm using this topology.
\end{enumerate}

As for step 2), it is important to see the advantage of being able to use different model types for the identified heterogeneous subsystems. In that way, problem-specific models (and in consequence also optimization procedures) can be used. In this research, the topology information connects these independent models as a more general interface for reunification and thus allows to find the optimal set-point for the overall system instead of just combining the local optima of the subsystems. However, we do restrict the model types to be quasi-static. That means we assume locally stabilized subsystems, so that all dynamics should occur in a shorter time interval than the one considered for modeling and optimization. Thus, we can abstract from these dynamics by averaging or alike methodologies. We do not consider the basal control technologies but want to determine set-points later used to guide these faster processes. In the work on hand we identified polynomials for all the dimensions of the subsystems as static energy models according to the following equation:
\begin{equation}
    f \left( \vec{\varphi}, \vec{\vartheta}, \vec{\xi} \right) = \left( \vec{\alpha}, \vec{\nu}, \vec{\chi} \right) , 
    \label{eq:model_equation}
\end{equation}
where $f$ is a static, potentially non-smooth function. $\alpha$ is the effort, in our example energy input, $\nu$ denotes the benefit, for our system the utilized energy output and $\varphi$ are the free parameters, which build the adjustable set-points of the subsystem. On the left side of the equation we take into account, besides the free parameters, external parameters $\vartheta$, considered in constraints and arbitrary but fixed parameters $\xi$, denoting parameters in which subsystems have to match, but for which there are no further requirements for their concrete values. On the right side we are mainly interested in efforts $\alpha$ and benefits $\nu$, but we can also define internal variables $\chi$ to be observed. The energy efficiency results in $\epsilon = \frac{\nu}{\alpha}$.
As described in step 1), to use topology information for optimization of an interconnected system of subsystems, this topology information has to be coded in an algorithmically tractable way. The presented approach suggests to formalize the decomposition in subsystems and topology as a directed graph. That means that the subsystem models as in Eq. \ref{eq:model_equation} get associated to one of the subsystems which together form the set of nodes of the graph. In the here discussed example we consider solely energy flows as interconnecting topology and integrate them in our system as the edges of the graph. The approach though is transferable to any kind of topology, like different kinds of material flows. The mathematical connection of the models according to their topology is realized by the formulation of constraints over the respective dimensions: If the benefit $\nu_A$ of subcomponent A acts as effort $\alpha_B$ for subcomponent B, then their values have to be (approximately) equal. These dimensions as parts of a system of EnPI and the hereon based modeling of energy-converting industrial systems is also elaborated in \cite{ThieleGregor2019Ffee}, \cite{optTopoCodit2020} and \cite{optTopoCodit2022}.

\subsection{Optimization Algorithm based on Topology}
The actual optimization is executed by an exemplary implementation of an algorithm based on topology (OptTopo). In this subsection, its principle of operation is presented. Due to the format of this paper, it can not be discussed in detail. For further insight, the papers \cite{optTopoCodit2020} and \cite{optTopoCodit2022} are suggested. The idea of solving a complex problem by means of a decomposition in smaller subproblems comes from the algorithmic paradigms \textit{Divide and Conquer} and \textit{Dynamic Programming}. We lean especially on the core idea of \textit{Dynamic Programming}, known as \textit{Bellman's principle}, that the optimal solution for the overall system must also be optimal for the tail-problems at any point in our procedure. In our case, the "tail" from any point on is given by the direction of the flow through the directed graph.

\paragraph{Optimization Problem}
The encoding of the topology knowledge as directed graph results in a structure which has a common root node (the virtual source of all energy consumed by the system) which forms the starting point of the system for OptTopo. To avoid the difficulties of a multi-objective optimization problem different kinds of energy input (e.g. gas and electricity) are made commensurable by an according weight (e.g. monetary equivalent) and merged in the common root node. We assume an acyclic system and graph so that all edges (energy flows) direct themselves towards a common sink (the benefit requested from the system). The optimization problem is now defined as follows: find, if possible, a configuration of the free variables of all the subsystems such that the request (at the sink) is satisfied with minimal energy demand (at the root).

\paragraph{Boundedness and Discretization}
In its present state, OptTopo uses a brute force approach to solve the elementary models for each subsystem, meaning that, at each subcomponent, it computes the solution for every possible combination of relevant parameters and saves only the best ones for each possible benefit. As a consequence, we assume all dimensions to be contained in bounded intervals that we are able to discretize. These limitations can be avoided concerning the parameters of the model in algorithms using topology by not using a brute-force algorithm for solving the subcomponents. But the boundedness and discretization of the efforts and benefits is part of the here developed utilization of topology and therefore has to be guaranteed for the presented approach. The following paragraph is explaining how these properties are used for the topology-based optimization.

\paragraph{Optimization based on topology}
The topology information is encoded in the graph structure of the problem with energy flows as edges between individual subcomponents (nodes). OptTopo uses this information to solve the overall optimization problem by computing it \textit{from left to right} - which is the direction of the topology respectively the energy flows. This means OptTopo creates a topological ordering of the nodes of the graph and then computes the subcomponents in this order. This topological ordering simply follows the paths of the energy flows and if two components are in the same hierarchical layer (with no edge between them pointing to one or the other), they are randomly ordered. To find the optimum for the overall-system, the individual optimum of each subcomponent is computed for all its possible benefit values. With this ordering, the benefit of a component A is always computed before the associated effort of a succeeding component B, since A is a predecessor of B. When all predecessors of a component are computed then the different values for the effort of the considered component can be propagated along all predecessors up to the common root. This strict procedure gives commensurable efforts and allows to calculate the efficiency in a certain node referencing to a common source (e.g. primary energy or cost).

For example, the benefit of a chiller plant (cooling power in \si{\kilo\watt}) may be bounded to an interval $[80,100]$ and we determine a discretization of steps of $ \SI{5}{\kilo\watt}$. OptTopo then computes an optimal solution for any (feasible) of these $ \SI{5}{\kilo\watt}$-steps. With this procedure there can not be an optimum of a later component which is turned unfeasible by decisions of prior components.

Thus, no possible solutions for the subsequent, not-yet-computed components is discarded to early. We call a subsystem of a node the set of components which are, according to the topological ordering, arranged before this component. At every point of the graph traversal OptTopo has already computed the optimal solution for this subsystem in such a way, that one optimal solution exists for every possible combination of benefits of the components of the subsystem. As pointed out before, this sequential computation \textit{from left to right} by graph traversal would not be possible with infinite sets of possible values for the benefits, since the storage capability is of course bounded. But with further developments of OptTopo, especially data structures and procedures that are better optimized, the granularity of the discretization can be designed to be sufficiently fine.

\paragraph{Optimization of Decomposed Subproblems}
As described above, each individual subcomponent is optimized using the described model in Eq. \eqref{eq:model_equation} for any given value in the set of possible benefits of the regarded subsystem. If there exist more possible solutions for a possible benefit, the opportunity for optimization arises and OptTopo saves only the optimal solution per benefit. This optimum is found by simple comparison. As already pointed out in the previous paragraph, this component optimization, nevertheless leads to a global optimum for the overall system since only those solutions are objected to a comparison which do not differ in any of the following dimensions: benefits, parameters or other constraints also used in models of not-yet-computed components.

\paragraph{Solution Look-up}
At this point the graph traversal is complete, meaning that OptTopo computed all the components, the optimal configuration for any given request in the requested discretization can be received by a simple look-up. This is a big advantage compared to conventional optimizers which would have to solve an individual problem for every demanded request. OptTopo however can solve multiple requests with no additional computation required. 

\section{Implementation and Experiments}

For the sake of brevity, we will not present details of the implementation of the testbed developed to evaluate OptTopo and compare it to other optimizers. The interested reader can obtain this information from former publications on the presented approach \cite{optTopoCodit2020} and \cite{optTopoCodit2022}.

\subsection{Comparison and Benchmarking}
OptTopo is compared to three established optimizers, COBYLA and SDPEN, which are local optimizers, and the global optimizer COUENNE.
\paragraph{Sequential Derivative-free Penalty method (SDPEN)} Introduced by Liuzzi in 2010, SDPEN minimizes via line search a \textit{sequential penalty function} consisting of the objective function plus a penalty term penalizing constraint violations \cite{liuzzi_sequential_2010}. During iteration, the stepsize of the line search and the multiplicative parameter of the penalty term are driven to zero and infinity respectively. The line search can be performed for any function but convergence to a stationary point of the original problem is only guaranteed for continuously differentiable objective functions and constraint terms (the other condition being that in each accumulation point of the sequence generated by line search, there exists at least one feasible search direction  along which all constraints which are broken by the current iterate can be lowered with a simple gradient step).
\paragraph{Constrained Optimization by Linear Approximation (COBYLA)} Similar to the sequential penalty function of SDPEN, COBYLA 
employs a \textit{merit function} combining the objective function and the state constraints \cite{powell_direct_1994}, where the penalty parameter is again increased heuristically during iteration. In contrast to SDPEN, however, it is not the merit function that is minimized in each step. Rather, in each iteration a \textit{linear approximation} of the objective function is defined by linear interpolation on a non-degenerate simplex. The resulting linear polynomial is now optimized in a trust region defined by a specified radius around the current "best" point (in terms of the merit function). Which point is to be replaced by the new iterate is mainly determined by the need to receive in each iteration an acceptable non-degenerate simplex, i.e. one whose volume does not collapse to zero. In some steps (see \cite{powell_direct_1994}), a new iterate is not chosen by optimization of the linear polynomial, but with the sole purpose of improving acceptability of the simplex.
\paragraph{Convex Over and Under Envelopes for Nonlinear Estimation (COUENNE)} Part of the COIN-OR infrastructure for Operations Research software \cite{belotti_branching_2009}, COUENNE deploys a \textit{spatial branch\&bound (sBB)} scheme to perform global optimization on Mixed Integer Nonlinear Programming problems (MINLP). In sBB, the initial problem is successively partitioned according to an sBB tree structure into smaller problems, each of which restricted to a subset of the set of possible solutions defined by the constraints. For these subproblems, lower bounds are obtained by convex relaxation. COUENNE first reformulates the original problem by the introduction of auxiliary variables, such that it is easier to obtain lower bounds. For the branch\&bound framework, Linear Programming relaxations are used. Critically, COUENNE provides several options to enhance performance such as bound tightening (reducing the size of the solution set) and branching strategies (minimizing the size of the sBB tree).

\subsection{Experimental Setup}

In this section we present the experiments conducted to evaluate the functional model of OptTopo. We designed three different experiments: In the first experiment in Sec. \ref{sec:feasible} the general feasibility of the approach is demonstrated and in Sec. \ref{sec:comparison} the quality of its solutions is compared to the comparison and benchmarking algorithms COBYLA, SDPEN and COUENNE. The last experiment in Sec. \ref{sec:scalability} examines the effect of a variation of the problem size on the optimizer.

The objective of the experimental optimization is to minimize the effort needed to deliver a specified cooling power to a consumer. The cooling complex comprises the systems and respective parameters listed in Tab. \ref{tab:groessenliste}. $\varphi_{\text{TempCoTo}}$ denotes the set-point temperature for the cooling power, ${\varphi_{\text{HysteresisA}}}$ and ${\varphi_{\text{HysteresisB}}}$ the hysteresis width of the two chillers and ${\varphi_{\text{SetPressureCoWa}}}$ and ${\varphi_{\text{SetPressureChWa}}}$ the set pressure for the Cooling water cycle and the Chilled water cycle respectively. The variable ${\xi_{\text{TempHeatex}}}$  is the temperature level between the cooling water cycle and the chillers. The temperature of the consumer $\xi_{\text{TempConsumer}}$ depends on both the unknown consumer and the cooling system considered. The polynomial models according to Eq. \eqref{eq:model_equation} for all the efforts and benefits of the subsystems of the decomposed system were identified based on a data set gained from a Simscape\textsuperscript{TM} simulation of this complex system. Besides these models the data set provided the occurring values for all the dimensions which were used to define bounded intervals on any dimension.

In Tab. \ref{tab:groessenliste} we can observe how the model of the given industrial system could be reduced by decomposition. For the decomposed system, the greatest amount of dimensions results for the modeling of the chillers, depending on 4 parameters. Since the subcomponents are computed independently and sequentially, the dimension for the decomposed system equals the maximal dimension of a subcomponent, which is 4 in this example. An analogous modeling of the overall system would depend on at least 5 dimensions, the free parameters $\Phi$. The problem size could hence be reduced by at least 1 dimension.

\begin{table*}
\centering
\caption{\textbf{Overview of variables for each subsystem.} While optimizing the entire system evokes a problem with five free influence variables, decomposing the entire system reduces to a maximum number of four dimensions to be considered from free variables $\Phi$ and arbitrary but fixed variables $\Xi$.}
\label{tab:groessenliste}
\begin{tabular}{|p{4mm}|*4{p{19mm}|}p{3mm}|}
\hline
 & Cooling tower & Cooling water & Chillers & Chilled water & $\Sigma$ \\ \hline
$\Phi$	& $\varphi_{\text{TempCoTo}}$ & ${\varphi_{\text{SetPressureCoWa}}}$&${\varphi_{\text{HysteresisA}}}$, ${\varphi_{\text{HysteresisB}}}$& ${\varphi_{\text{SetPressureChWa}}}$ & \textbf{5} \\ \hline
$\Xi$	&	${\xi_{\text{TempHeatex}}}$ 	& ${\xi_{\text{TempHeatex}}}$ & $\xi_{\text{TempHeatex}}$, ${\xi_{\text{TempConsumer}}}$  & $\xi_{\text{TempConsumer}}$ & 2 \\ \hline
$\Sigma$ &  2	& 2	&\textbf{4} & 2& \\ \hline
\end{tabular}
\end{table*}

\subsection{Results and Interpretation}

\subsubsection{Feasibility of the Approach}
\label{sec:feasible}
\begin{figure}
    \centering
    \includegraphics[width=\linewidth]{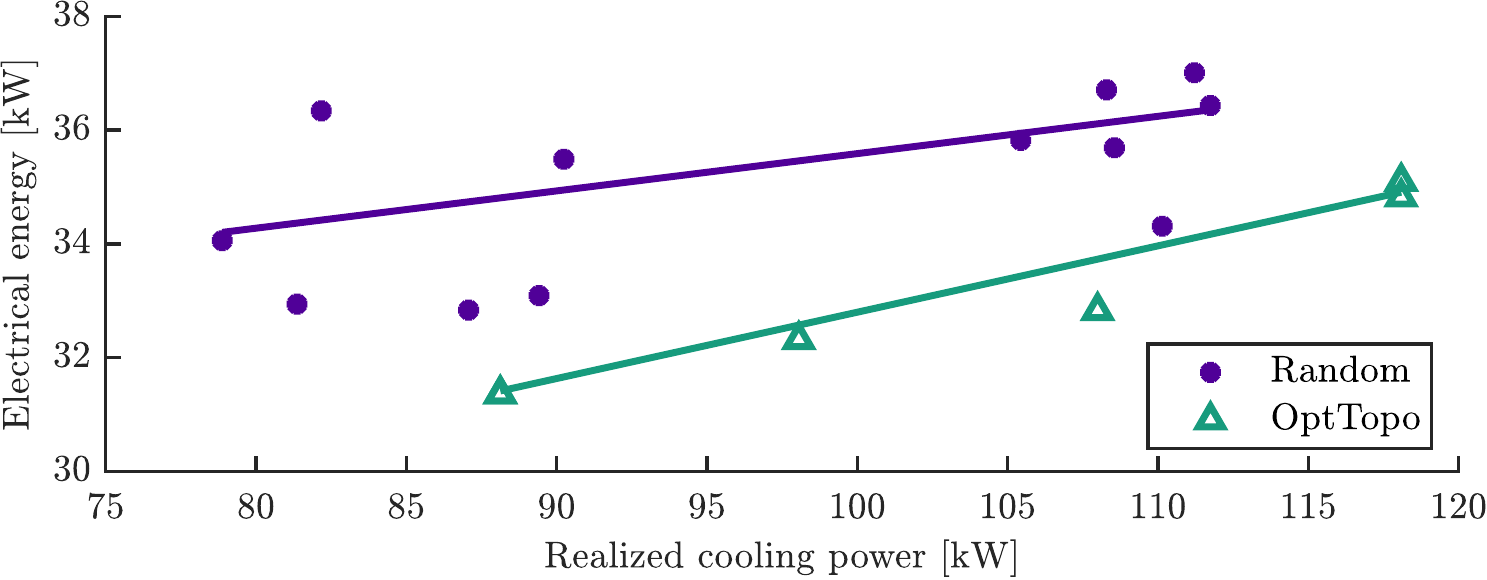}
    \caption{Operation points resulting from solutions from OptTopo together with operation points from randomly chosen configurations. The former ones prove to realize a lower demand of electrical energy in this sample. A quadratic regression illustrates this trend.}
    \label{fig:N1_MaerzCompareToRandom_ieee}
\end{figure}

This experiment shows the general functionality of the approach and the usability of the generated solutions. OptTopo was configured to optimized the described cooling system for 5 different requests equally distributed over the interval from 74 to 121 kW. Afterwards the solutions are realized using the simulation of the system to demonstrate the feasibility and to compare their efficiency with randomly generated points. For this configuration OptTopo ran approximately 1 hour and called 10 bn functions. There was no solution found for the lowest request. For the last request however, OptTopo returned two solutions. This can occur if the predicted effort is (approximately) equal, in which case the selection of the optimal configuration should be based on a different operation criterion. Fig. \ref{fig:N1_MaerzCompareToRandom_ieee} illustrates that OptTopo achieves valid set-points which lead to more efficient operating-points in comparison to random settings.

\subsubsection{Comparison to Other Optimization Procedures}
\label{sec:comparison}
To evaluate the performance of the here presented algorithm OptTopo and the efficiency of its returned solutions beyond its general feasibility, the following experiment compares it to the comparison and benchmark algorithms COBYLA, SDPEN and COUENNE. The amount and the interval of requests posed are the same as in the first experiment listed in the previous Sec. \ref{sec:feasible}. 
For the lowest request, the other optimizers too fail to find a solution, hinting to an error in the model which renders this point infeasible. In addition, COBYLA does not find a solution for the second request. SDPEN and COUENNE, like OptTopo, find a solution for all the remaining 4 requests.
\begin{figure}
    \centering
    \includegraphics[width=\linewidth]{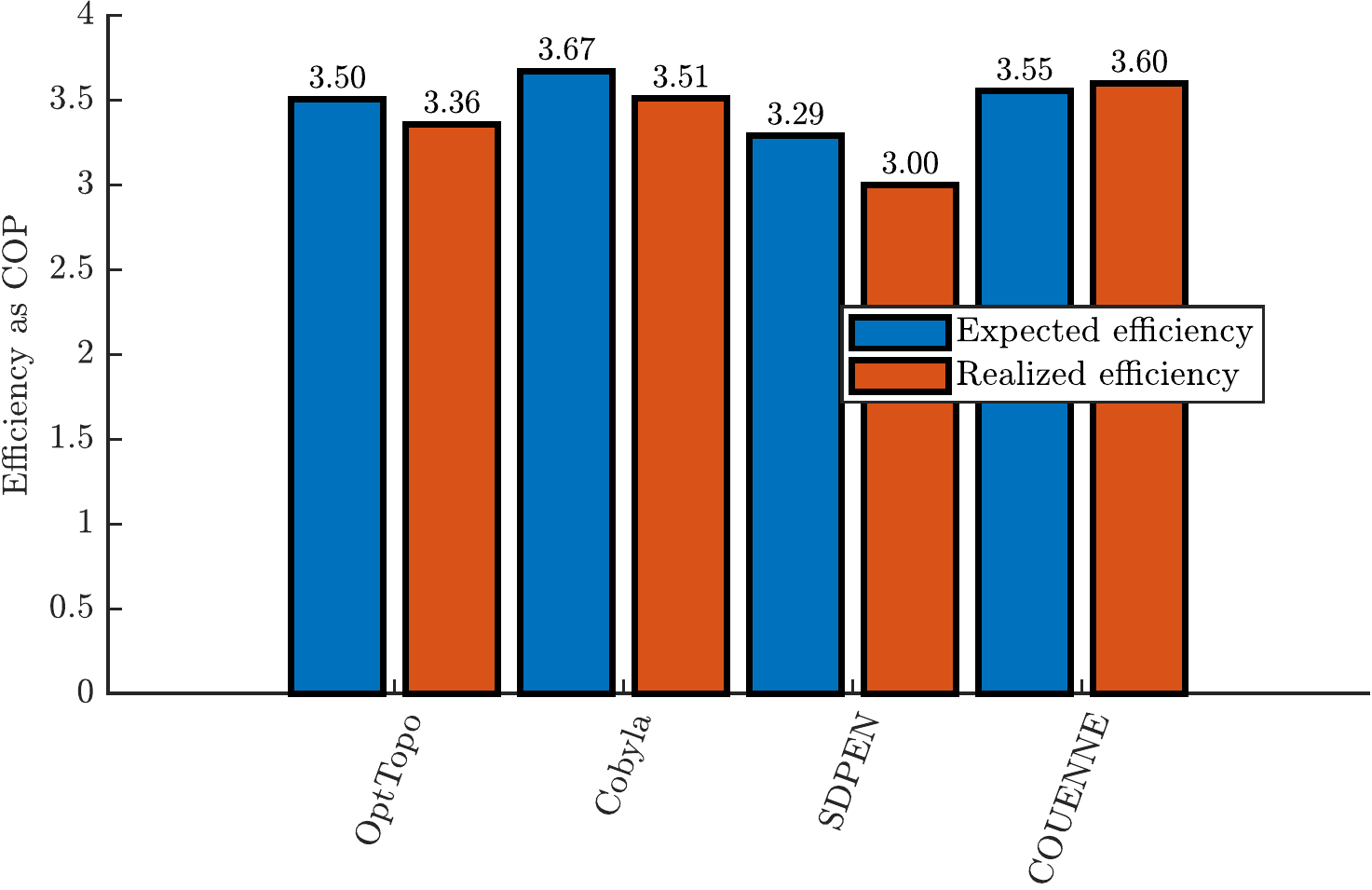}
    \caption{Efficiency of the optimizers in comparison}
    \label{fig:N1_MaerzEfficiency_ieee}
\end{figure}
Fig. \ref{fig:N1_MaerzEfficiency_ieee} shows the expected efficiency (blue, computed by the optimizer) and the realized efficiency (red, running the simulation with the returned solution) for one exemplary chosen request for all the optimizers. As expected, the global solver COUENNE reaches the best value. It is followed by COBYLA and OptTopo, while SDPEN shows the poorest efficiency. Notably, SDPEN also displays the largest discrepancy between expected and realized efficiency. For COUENNE there is only a small difference between these values (being the only algorithm which underestimates the efficiency of its solution) while for COBYLA and OptTopo both slightly overestimate their efficiency. The good (realized) result of the benchmark algorithm COUENNE yields a strong case for the suitability of the identified polynomials. In terms of efficiency, the first prototype of OptTopo deployed here seems to be able to successfully compete with the other local solvers COBYLA and SDPEN.

\begin{figure}
    \centering
    \includegraphics[width=\linewidth]{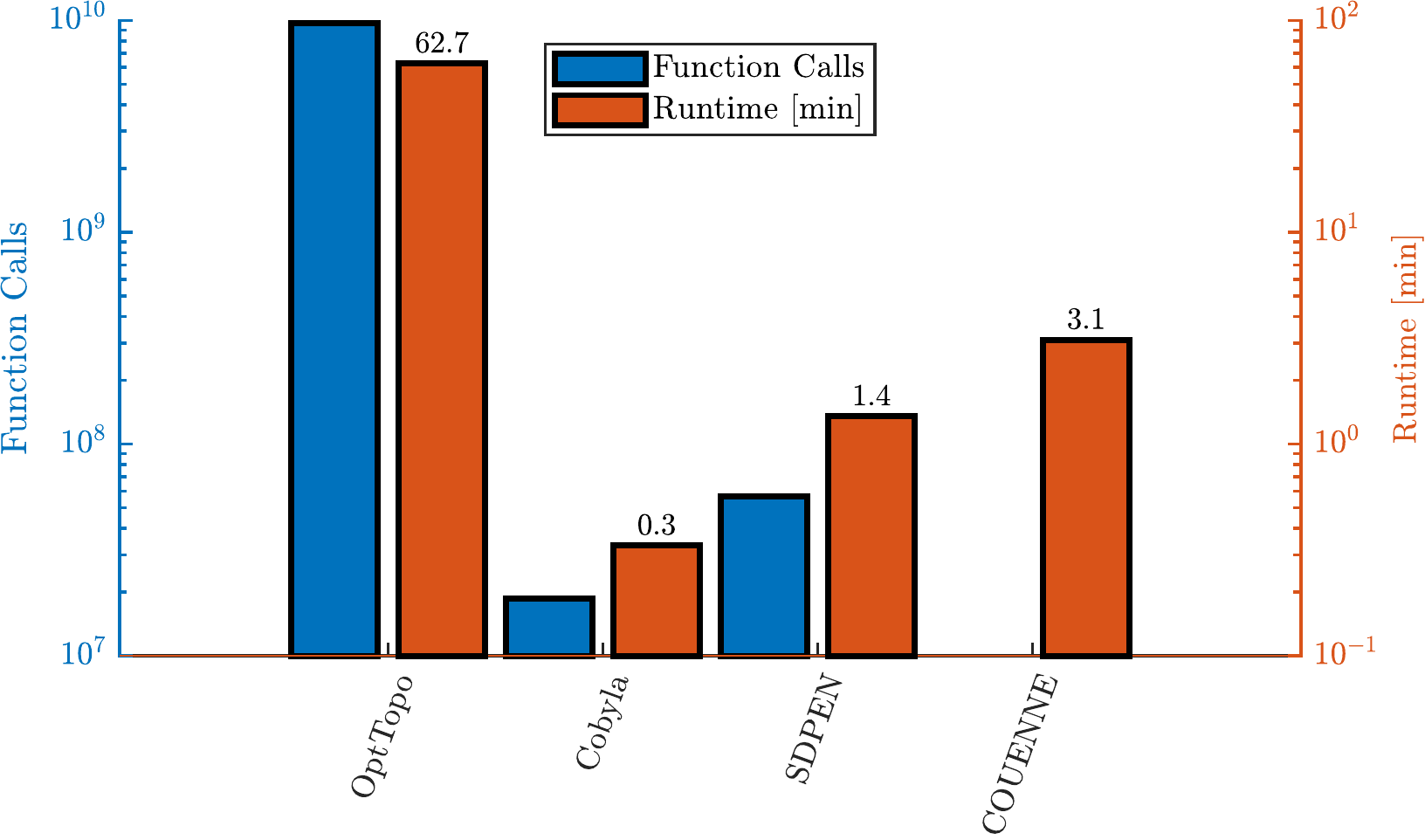}
    \caption{Runtime of the optimizers in comparison}
    \label{fig:N1_MaerzRuntime_ieee}
\end{figure}

Fig. \ref{fig:N1_MaerzRuntime_ieee} compares the runtime and function calls of this experiment for all the optimizers. Clearly, OptTopo is much more expensive than the other approaches. This high complexity is caused by the brute force-approach used to solve the models for the subcomponents, which leads to a comparatively large number of function calls. As already stated in Sec. \ref{sec:concept}, this needs to be optimized via better suited data structutes and methodologies like parallelization not yet realized in this first prototype version of the algorithm.

\subsubsection{Scalibility}
\label{sec:scalability}
\begin{figure}
    \centering
    \includegraphics[width=\linewidth]{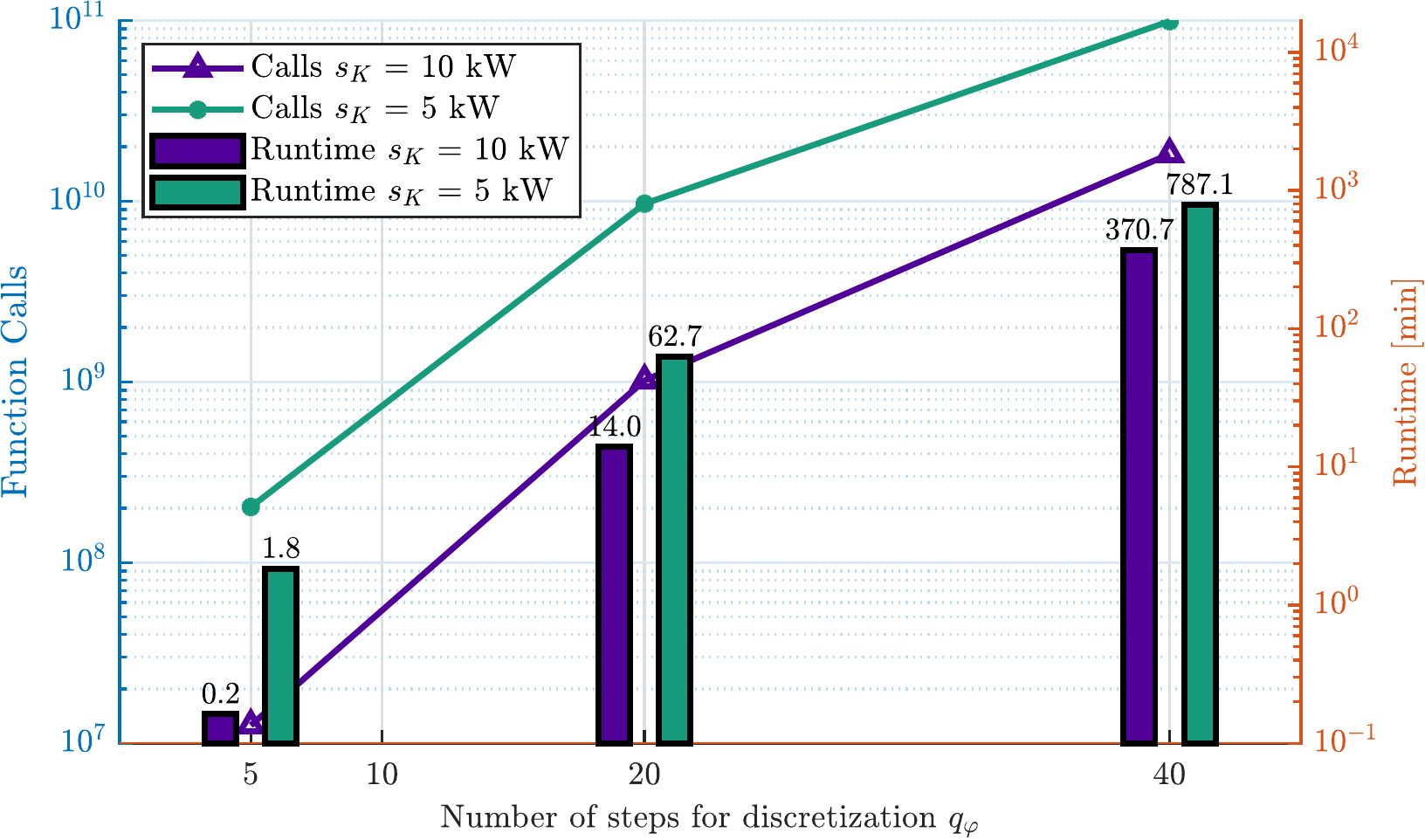}
    \caption{Runtime of OptTopo for different numbers of discretization steps $q_\varphi$}
    \label{fig:f9_MaerzProblemIncreaseRuntime_ieee}
\end{figure}

In this last experiment, the scalability of the approach is examined. In the current implementation, OptTopo uses a brute-force algorithm to compute the optimal solutions of the subsystems. Doing so, the parameters of the modelfunction have to be discretized in a suitable manner and OptTopo is based on a fixed grid with $\Phi$ and $\Xi$ as independent dimensions. All these dimensions are discretized in $q_\varphi$ steps. Besides that, the energy flows need to be discretized as described in Sec. \ref{sec:concept}. For this discretization we chose to run only two different settings: A coarse discretization and a finer one. The flow of cooling power is discretized in steps of $s_K = \SI{5}{\kilo\watt}$ or $s_K =\SI{10}{\kilo\watt}$. This setting implies for  $s_K = \SI{5}{\kilo\watt}$ a fine discretization of the electrical power in \SI{1}{\kilo\watt} steps and for $s_K = \SI{10}{\kilo\watt}$ a coarser discretization of the electrical power in \SI{2}{\kilo\watt} steps. For these two settings, Fig. \ref{fig:f9_MaerzProblemIncreaseRuntime_ieee} shows the effect on the runtime of a configuration with different values for the granuality of the discretization of the free variables with $q_\varphi = \numlist{5;20;40}$. 

As expected, both runtime and number of function calls grow with a finer discretization for the energy flows (purple versus green columns) and also with a finer discretization of the parameters (x-axis of the plot). This relation is also displayed in Tab.  \ref{tab:RuntimeN1_N2}: The grid has a size of $q_\varphi^{\lvert \Phi \cap \Xi \rvert}$ since all dimensions considered are divided into $q_\varphi$ steps. In the presented use-case the maximal number of dimensions is given by $\lvert \Phi  \cap \Xi \rvert = 4$ for the chillers (see Tab. \ref{tab:groessenliste}). Thus, for $\tilde{q}_\varphi=2\cdot q_\varphi$, the size of the grid grows to $\tilde{q}_\varphi^{\lvert \Phi \cap \Xi \rvert}$ by a factor $2^4=16$. Tab. \ref{tab:RuntimeN1_N2} shows that  the number of function calls increases by factor 10 instead of 16 when changing from \num{20} to $40$ steps, and when changing from $5$ to $20$ steps, the amount of combinations grows by factor 256 but the amount of function calls only by a factor of approximately 50. This effect results from the component-wise optimization that allows to discard infeasible and not efficient solutions for equal benefits at an early stage of the graph traversal so that not all of the possible combinations have to be saved.

\begin{table}
\centering
\caption[Runtime of OptTopo]{\textbf{Runtime of OptTopo depending on the discretization.} 
The computational effort increases with the number of steps in the grid but this increase is slower than the expected exponential growth.}\label{tab:RuntimeN1_N2}
\begin{tabular}{|c|c|c|c|}
\hline
	&  \textbf{Combinations}&\textbf{Function calls} & \textbf{Runtime} in \si{\minute}\\ \hline \hline
	5  Steps & \num{625} & \num{2.03e8} & \num{1.8395} \\ \hline
	Factor   & \num{256} & \num{47.6512} & \num{34.0853} \\ \hline
	20 Steps & \num{16e4} & \ \num{9.6732e09} & \num{62.7} \\\hline
	Factor 		& \num{16} & \num{10.2292} & \num{12.5534} \\ \hline
	40 Steps & \num{256e5} & \num{9.8949e10} & \num{787.1} \\ \hline
\end{tabular}
\end{table}

\section{Conclusion and Outlook}

We presented a set-point optimization algorithm, OptTopo, which utilizes topological information of a complex system consisting of several subsystems connected by directed energy flows. We demonstrated OptTopos performance in comparison to widely used local solvers SDPEN und COBYLA as well as the global solver COUENNE. While the presented prototype is prone to high complexity, due to the brute-force approach of the subsystem optimization, it could be demonstrated to successfully compete with the aforementioned local solvers in terms of reliability and efficiency. Additionally, once the algorithm terminates once, an optimal configuration for any given request adhering to the
requested discretization can be received by a simple look-up, generating no additional cost. This is a big advantage compared to other widely used optimizers.

The mitigation of OptTopo's complexity is a conjecture for future work. In its present form, the algorithm does not use parallelization, which could greatly increase the speed of optimization in systems with multiple parallel branches.
The present version also requires free and internal parameters to be contained in finite intervals, which are then discretized for the brute-force optimization of the subsystems, leading to an excess of function calls compared to other methods. More sophisticated continuous optimization methods for the subproblems, even hybrid versions with local solvers such as SDPEN or COBYLA could be introduced for the subsystems in order to reduce computational burden.

\section*{Acknowledgment}

We gratefully thank Rico Clau\ss\ from Technical University Berlin for his work on the concept, the implementation and early experiments of the approach presented. Furthermore, we would like to thank Knut Grabowski from \textit{ÖKOTEC Energiemanagement GmbH} for his support regarding the EnPI methodology, his impulses to the optimization method and further helpful discussions. We thank Rosa Bartholdy for fruitful discussion on the presentation of the concept. 

\bibliographystyle{ieeetr}
\bibliography{Literature.bib}

\end{document}